# Probing Magnetism in Insulating $Cr_2Ge_2Te_6$ by Induced Anomalous Hall Effect in Pt

Mark Lohmann[1], Tang Su[2,1], Ben Niu[3,1], Yusheng Hou[4], Mohammed Alghamdi[1], Mohammed Aldosary[1], Wenyu Xing[2], Jiangnan Zhong[2], Shuang Jia[2], Wei Han[2], Ruqian Wu[4], Yong-Tao Cui[1], and Jing Shi[1]

1. Department of Physics and Astronomy, University of California, Riverside, California 92521, USA
2. International Center for Quantum Materials, School of Physics, Peking University, Beijing 100871, P. R. China
3. National Laboratory of Solid State Microstructures, Department of Materials Science and Engineering, College of Engineering and Applied Sciences, and Collaborative Innovation Center of Advanced Microstructures, Nanjing University, Nanjing 210093, China
4. Department of Physics and Astronomy, University of California, Irvine, California 92697, USA

*Correspondence to: jing.shi@ucr.edu.

## Abstract

Two-dimensional ferromagnet $Cr_2Ge_2Te_6$ (CGT) is so resistive below its Curie temperature that probing its magnetism by electrical transport becomes extremely difficult. By forming heterostructures with Pt, however, we observe clear anomalous Hall effect (AHE) in 5 nm thick Pt deposited on thin (< 50 nm) exfoliated flakes of CGT. The AHE hysteresis loops persist to ~ 60 K, which matches well to the Curie temperature of CGT obtained from the bulk magnetization measurements. The slanted AHE loops with a narrow opening indicate magnetic domain formation, which is confirmed by low-temperature magnetic force microscopy (MFM) imaging. These results clearly demonstrate that CGT imprints its magnetization in the AHE signal of the Pt layer. Density functional theory calculations of CGT/Pt heterostructures suggest that the induced ferromagnetism in Pt may be primarily responsible for the observed AHE. Our results establish a powerful way of investigating magnetism in 2D insulating ferromagnets which can potentially work for monolayer devices.

Understanding the magnetic properties of two-dimensional (2D) van der Waals ferromagnets such as $Cr_2Ge_2Te_6$(CGT), $CrI_3$, amongst a plethora of others, has attracted a great deal of interest due to their ability to be exfoliated down to the monolayer allowing for studying magnetism in 2D systems[1,2] as well as their ability to easily form heterostructures with other 2D materials such as graphene and a wide range of transition metal dichalcogenides[3]. In $CrI_3$, for example, the intriguing interlayer antiferromagnetic coupling is uncovered when a few layers are involved. Recently, large tunneling magnetoresistance between the atomic layers in $CrI_3$ has revealed rich spin states as a result of the antiferromagnetic coupling[4]. To date, no tunneling experiment has been reported in the other 2D insulating ferromagnet, CGT, to study the magnetic properties such as the interlayer coupling. In conducting ferromagnets, transport measurements such as magnetoresistance and anomalous Hall effect (AHE) are routinely employed to probe the magnetic properties especially in small devices[5,6]. In insulating ferromagnets, however, such transport measurements are not directly applicable. In this study, we explore an alternative way to probe the magnetic properties of thin CGT flakes by electrical means. In principle, this method can be extended to devices made of monolayers of 2D magnets.

Although bulk CGT crystal is conductive at high temperatures, its resistance becomes extremely high in the ferromagnetic phase below 60 K[7–9]. Under very large bias voltages, the two-terminal resistance and magnetoresistance of CGT crystal could be measured[10]. Large back gate electric field was used to lessen the insulating behavior of thin CGT flakes (~10-50 nm) so that the AHE could be detected[11]. In recent studies of three-dimensional (3D) magnetic insulators such as $Y_3Fe_5O_{12}$ and $Tm_3Fe_5O_{12}$, an effective way of probing their magnetic properties is to take advantage of the induced transport properties in a metal layer such as Pd, Pt, Ta, etc. by forming heterostructures with the former[12–16]. The magnetic insulator properties are thus imprinted in the induced transport properties such as magnetoresistance and AHE in the metal layer. Similar proximity induced polarization-dependent photoluminescence was found in $WSe_2$ by forming heterostructure with $CrI_3$[3]. Here we adopt the same approach by fabricating CGT/Pt heterostructures with exfoliated CGT flakes and detecting induced magneto-transport properties in Pt.

Before studying induced magneto-transport properties, we first characterize the intrinsic properties of CGT itself. Magnetization measurements are performed on bulk CGT crystals using a SQUID magnetometer. The Curie temperature, $T_c$ ~ 61 K, was determined from the abrupt drop in magnetization in the temperature dependence from 5 K – 300 K when a 1 kOe magnetic field is applied parallel to the a-b plane of the crystal[9,10]. Magnetization measurements below $T_c$ with an external magnetic field either parallel or perpendicular to the a-b plane reveal soft ferromagnetic behavior with perpendicular magnetic anisotropy determined by comparison of the field required to saturate all the spins in both geometries, i.e., $H^{\parallel}_{sat} > H^{\perp}_{sat}$[9,10]. To confirm the transport behavior in thin exfoliated CGT flakes, we improved the Hall bar device fabrication to minimize the contact resistance by avoiding lithographic process after the exfoliation. We first patterned 5 nm Pt electrodes onto an exfoliated flake of boron nitride (BN) on a $SiO_2$ substrate by electron beam lithography (EBL). A second flake of BN was used to pick up and transfer a CGT flake onto the pre-patterned Pt electrodes. The top BN is larger than the CGT so that the two BN flakes encapsulate the CGT flake as is illustrated in Figure 1a. All the transfer and exfoliation of CGT was performed in an argon filled glovebox with < 0.1 ppm $H_2O$ and $O_2$ to protect the flakes from degradation. After transfer, no further fabrication steps took place and the device was exposed to ambient conditions for < 20 minutes during the device mounting before it was moved into an evacuated cryostat. We measured the transport properties of the device from 300 K down to 5 K through the I-V characteristics. Figure 1c shows the current measured while sweeping bias voltage between the source and drain electrodes for representative temperatures between 5 K and 100 K. In the range of 5 K – 60 K, the measured current is negligibly small for $|V_{Bias}| < 5$ V. Inset shows the I-V curves obtained at higher temperatures, 200 K and 300 K, both well above the magnetic ordering temperature with vertical scale in µA. Over the temperature range of 100 – 300 K, the contact resistance is relatively small. We calculated the resistance in the linear I-V regime over this temperature range and determined the band gap, $E_G$ ~ 0.23 eV, by fitting the thermal activation model, $\rho(T) = \rho_0 \exp(\frac{E_G}{2k_B T})$, as shown in the inset of Figure 1d. This value is similar to the reported value of $E_G$ ~ 0.2 eV[9,10]. Because of the very insulating behavior at and below $T_c$, it is very difficult to observe any AHE signal in the ferromagnetic phase.

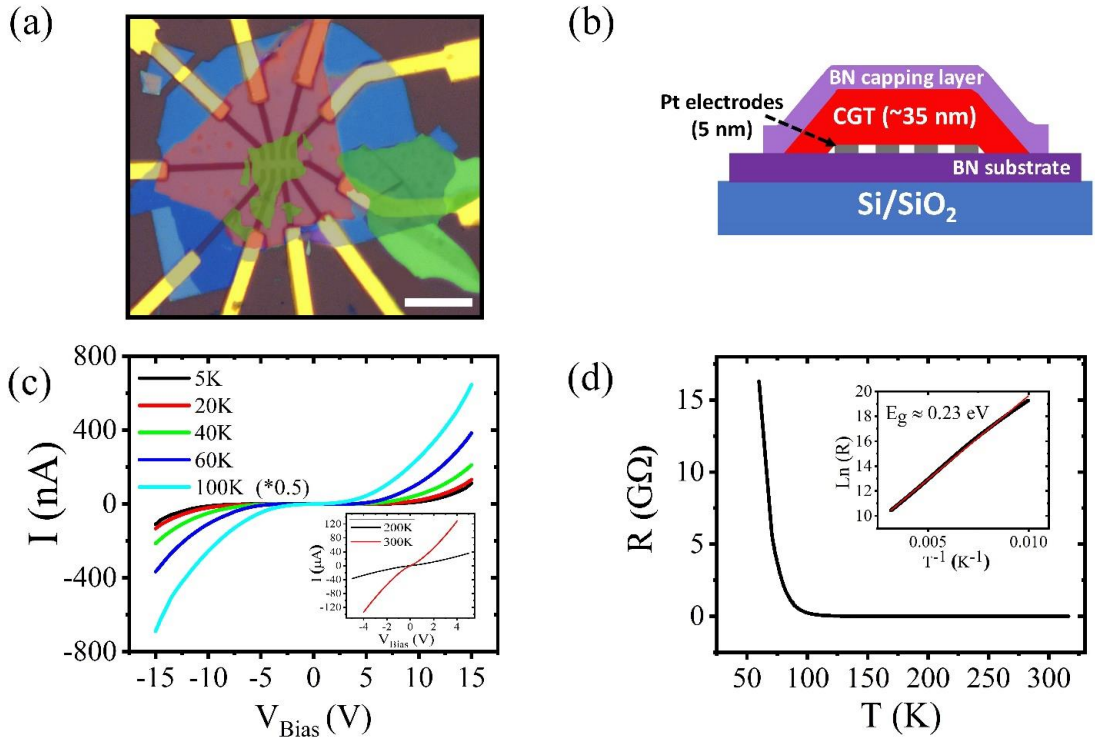

**Figure 1.** Transport properties of CGT device. (a) Optical micrograph of CGT transport device with 10 μm scale bar. False coloring is used to distinguish the different material layers: bottom BN in blue, CGT in green, and top BN in red. (b) Schematic illustration of the side view of the device in (a). (c) I-V characteristics of device in (a) measured at representative temperatures down to 5 K with 200 K and 300 K inset with vertical scale in μA. (d) Resistance vs. temperature measurement for the temperature range of 320 K to 60 K. Inset: Logarithmic plot of resistance vs. 1/T over the temperature range of 100-300 K with linear fit to extract the band gap.

To have strong induced magneto-transport properties in Pt, it is important to have a very clean interface in the CGT/Pt heterostructures[17]. Due to the air sensitivity of CGT flakes (more information on the degradation of CGT in ambient conditions can be found in Supporting Information), we also modified the standard fabrication steps for less air sensitive 2D materials to minimize the interface degradation. A schematic of this fabrication process is given in Figure 2a and further details in the Methods section. In the existing setup, we could not transfer samples from the glovebox into the sputtering chamber without exposing to air; therefore, we performed the exfoliation in the load lock of the sputtering system, and immediately evacuated the chamber to greatly reduce the $O_2$ and $H_2O$ exposure time to the freshly cleaved CGT surface, followed by Pt deposition. The Pt layer is not only the active layer for sensing the induced AHE, but also serves as a capping layer to prevent further oxidation in the subsequent device fabrication. Once removed from the sputtering chamber, the sample with many exfoliated flakes was viewed under optical microscope to locate a desired piece for making device ((i) in Fig. 2a). EBL and lift-off process were performed to deposit 60 nm thick Au electrodes ((ii) in Fig. 2a). Inductively coupled plasma etching was then used to etch the Pt layer to form Hall bar structure ((iii) and (iv) in Fig. 2a). Figures 2b and 2c show a representative device on a ~ 35 nm thick CGT flake and the height profile of the same device obtained with atomic force microscopy (AFM). The fabrication steps adopted here allow for a high device yield and are critical to achieving a high-quality CGT/Pt interface enabling the AHE in all devices we studied.

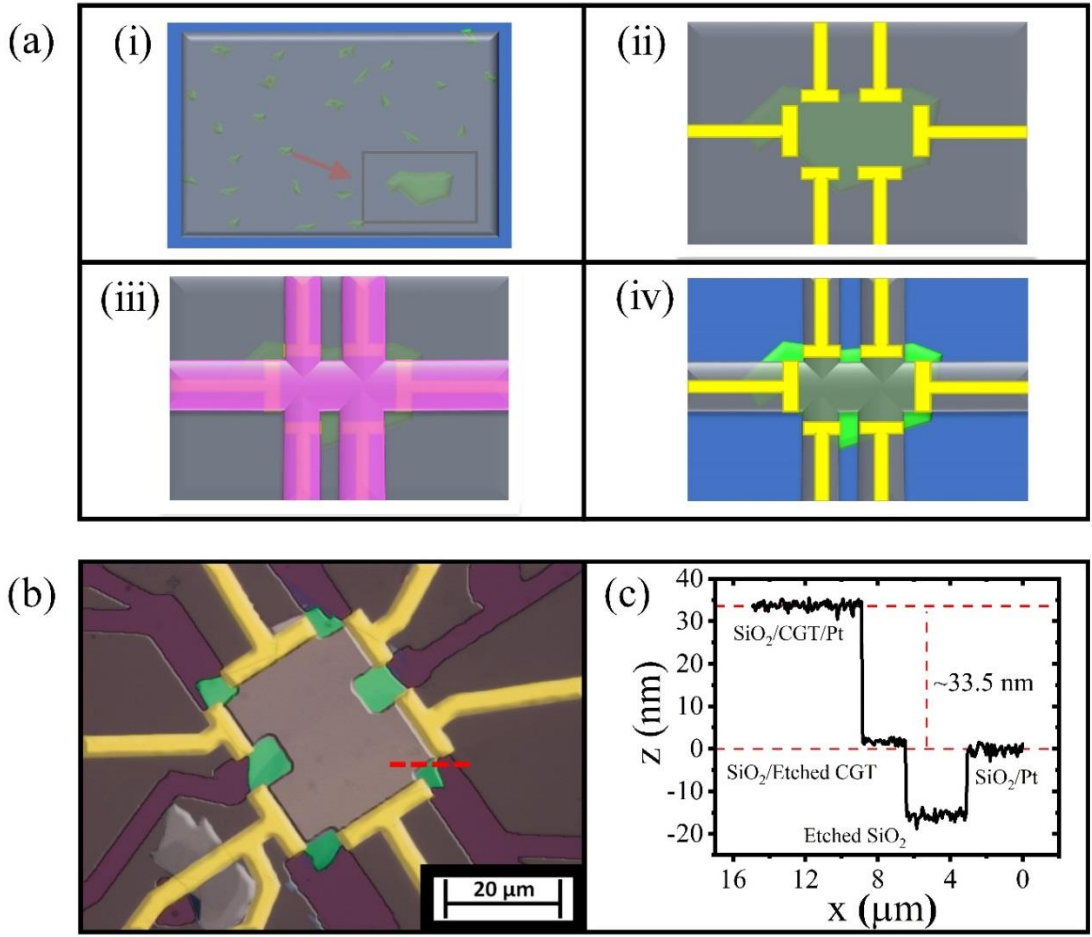

**Figure 2.** The fabrication of CGT/Pt hybrid devices. (a) Schematic of the device fabrication process. (i) Exfoliated CGT flakes covered by 5 nm Pt. (ii) Au electrode deposition on the chosen CGT flake shown in the box in (i). (iii) E-beam resist mask (cyan) for defining a Hall bar. (iv) Pt etching and e-beam resist removal. (b) Optical micrograph of CGT/Pt heterostructure with false color to clarify different regions of the device. False coloring is used to distinguish the different material layers: Pt in gray, CGT in green, Au in yellow, and $SiO_2$ in dark red. (c) Line cut from AFM image of CGT/Pt device. The cut line is represented by the red dashed line in (b).

Using this method, we successfully fabricated multiple CGT/Pt heterostructure devices in a range of CGT thickness down to ~ 35 nm. Thinner exfoliated CGT flakes are typically too small to be fabricated into the Hall bar device structure. The data presented in this main text all came from the device pictured in Figure 2b. Results from some of the other devices are qualitatively similar and are included in Supporting Information. Figure 3a shows the induced AHE data in Pt measured at selected temperatures from 5 to 65 K using a current of 2.0 mA in applied magnetic fields perpendicular to the cleaved plane, i.e., the 2D layers. These hysteresis loops are obtained after subtracting a linear ordinary Hall background (an example of full Hall hysteresis is shown in Fig. 4a) for all temperatures. Since Pt itself is paramagnetic and no hysteresis loop is expected for standalone Pt, the fact that AHE in Pt only appears below $T_c$ of CGT indicates that it is caused by the presence of the ferromagnetic CGT. Therefore, Pt AHE hysteresis merely reflects the magnetic hysteresis of the underlying CGT. In fact, similar slanted Kerr rotation hysteresis loops of ~ 19 nm CGT were recently reported[18] as well as in $Fe_3GeTe_2$ flakes >15 nm[19]. Note that the AHE hysteresis loops are significantly slanted and only a small fraction of the saturation value of AHE resistivity is retained at zero magnetic field. It suggests formation of non-uniform magnetization configurations such as domains. To obtain the spontaneous AHE resistivity, we extrapolate the high-field linear background to zero field and plot the intercept as a function of temperature in Figure 3b. Above 60 K, AHE signal falls below the noise level, and the only remaining Hall signal is from the ordinary Hall effect of the Pt layer. The magnetoresistance data at T = 4 K are also consistent with the slanted AHE hysteresis loop, as shown in Fig. 3c.

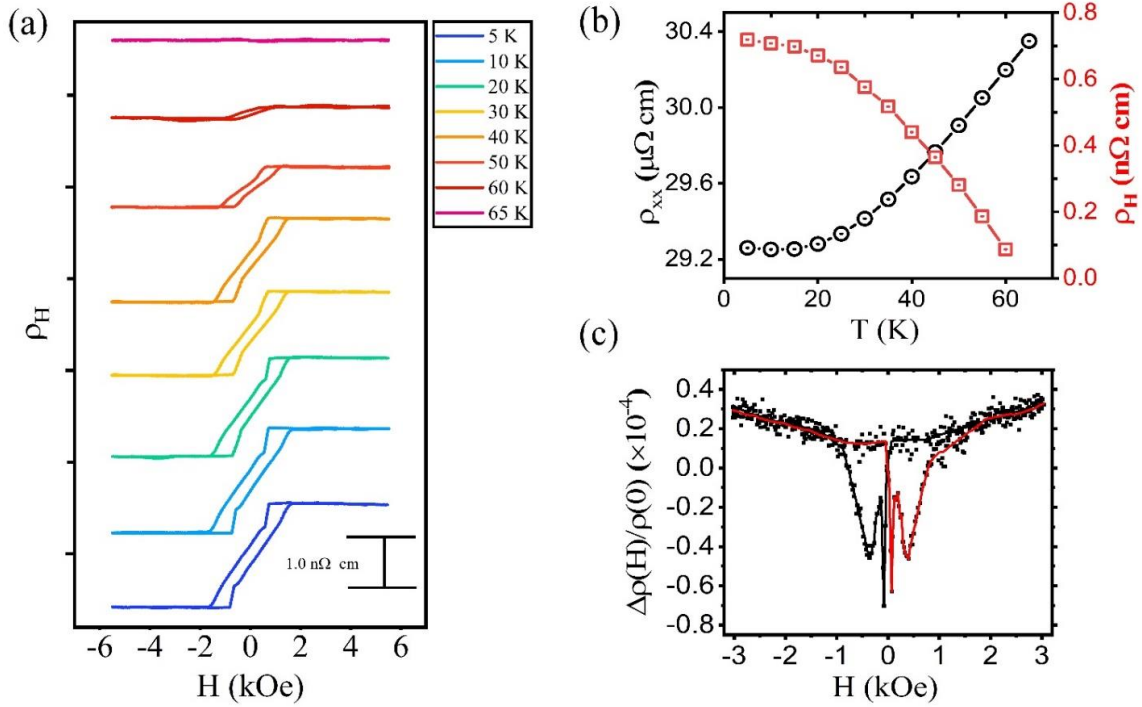

**Figure 3.** (a) AHE hysteresis loops for select temperatures from 65 K to 5 K after subtraction of the linear ordinary Hall background. (b) Magnitude of the measured longitudinal resistivity and Hall voltage as a function of temperature. (c) Magnetoresistance measured at 4 K.

It is known that the multi-domain or vortex state is favored in thick patterned ferromagnetic films[20] due to the dipolar energy winning over the exchange energy. To understand the Hall hysteresis loops of Pt which mirrors the magnetization state of CGT, we performed low-temperature magnetic force microscopy (MFM) measurements under applied magnetic fields. Along with a 4 K AHE loop of Pt, Figure 4 shows a series of MFM images taken on a different CGT/Pt device of a similar CGT thickness under perpendicular magnetic fields at 7 K. The applied magnetic fields during MFM imaging correspond to different points labeled in the hysteresis loop. In the present measurement geometry, the MFM signal characterizes the second order derivative of the out-of-plane component of the stray magnetic field, i.e., $d^2H_z/dz^2$. For samples with out-of-plane magnetization, MFM has responses both inside individual domains and at the domain walls. While it may not be straightforward to directly associate the MFM signal with the magnetization direction, the spatial variation of MFM signals qualitatively indicates the abundance of the domains which are seen in several images presented below. For a 2 kOe applied field in the z-direction (state 1), the magnetization of CGT is fully saturated, and it should be in a single-domain state. It is indeed confirmed by the uniform MFM contrast. As the field is lowered, the single-domain state is preserved until the field reaches ~ 0.5 kOe (state 2) at which point multiple domains emerge. It is in good correspondence to the reduced AHE signal from the saturation value. The sharp drop at point 2 in the AHE loop signals the domain nucleation. As the field is reduced further, the opposite domains expand, leading to stronger MFM contrast, and the AHE signal decreases accordingly. This trend continues until the field reaches -1.5 kOe at which field an oppositely oriented single-domain state is realized (state 4). Although the magnetization direction is reversed compared to state 1, the MFM contrast remains the same because the tip magnetization, having a relatively low coercive field of ~400 Oe, is also reversed thus generates a force gradient in the same direction as the opposite saturation field (state 1). When the magnetic field is reversed, a similar trend is observed, and similar domain nucleation and expansion patterns are displayed at points 4 and 5. This sequence of MFM measurements shows close correspondence to the AHE loop and thus confirms that the AHE signal in the Pt layer tracks the behavior of the underlying CGT flake.

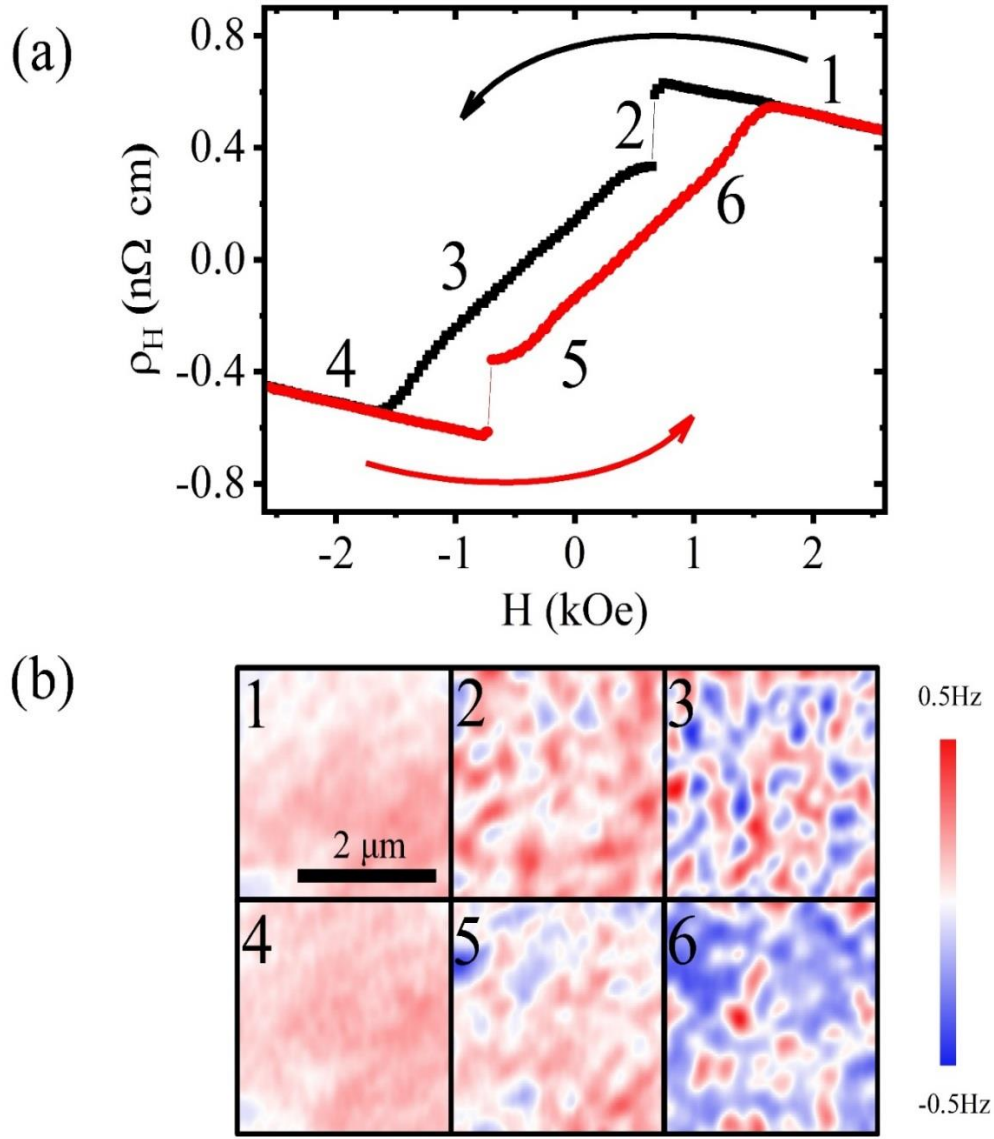

**Figure 4.** (a) Anomalous Hall hysteresis loop, measured at 4 K. (b) Frequency shift contrast images of CGT/Pt taken with low-temperature MFM at different magnetic fields that correspond with the points marked in (a).

In heterostructures containing 3D magnetic insulator and a heavy metal layer such as Pt, there is a debate about the mechanism of the induced AHE, i.e., whether it is due to induced magnetism in Pt or a spin current effect[10,20–22]. Either mechanism can imprint the magnetization states in the AHE of Pt. Although it is not the primary focus of this work, we explore the origin of the induced AHE in CGT/Pt heterostructures by performing density functional theory (DFT) calculations. Computational details of four atomic layers of Pt (4L-Pt) on CGT are shown in the Supporting Information. The real-space distribution of spin density ($\Delta\sigma = \rho_\uparrow - \rho_\downarrow$, where $\rho_\uparrow$ and $\rho_\downarrow$ are spin-up and spin down charge densities, respectively) and its planar average of the CGT/4L-Pt heterostructure are shown in Fig. 5(a) and (b). Clearly, ferromagnetic CGT introduces noticeable spin polarization in all Pt layers. Interestingly, $\Delta\sigma$ around Pt atoms in the first layer oscillates rapidly in the lateral plane, as they align differently with Cr atoms, and its planar average is small in the first Pt layer and maximizes in the second Pt layer. This is different from previous results for Pt films on traditional magnetic films such as Fe, Co and Ni, where the atomic alignment at the interface is much simpler[17]. Overall, the averaged induced magnetic moment in of CGT/4L-Pt is 0.0074 $\mu_B$ per Pt atom and parallel to the magnetic moments of $Cr^{3+}$ ions. To examine the range of induced spin polarization in Pt, we also calculated the CGT/22L-Pt (5 nm Pt, which is comparable to our experimental samples) heterostructure and found that the average magnetic moment drops to 0.0009 $\mu_B$ per Pt atom. Using the method proposed by Y. Yao et al[24], we determined the anomalous Hall conductivity (AHC) of CGT/4L-Pt heterostructure to be 2700 S/m. To estimate the AHC in CGT/Pt heterostructure with 5 nm thick Pt, we first studied the AHC in a toy model, i.e., AHC in monolayer Pt that is subjected to exchange interaction and consequently acquires a magnetic moment. By varying the exchange strength, we found that the AHC linearly decreases as the magnetic moment of Pt decreases from 0.01 $\mu_B$ per Pt atom to zero as shown in Figure 5c. This linear relation allows us to estimate the AHC value in thicker Pt films if the magnetic moment of the Pt films is calculated. In fact, the fully calculated AHC and induced magnetic moment for the 4L-Pt fall right on the extended straight line. We calculated the average magnetic moment of the CGT/5 nm-Pt heterostructure and obtained 0.0009 $\mu_B$ per Pt atom. Based on Figure 5c, we estimated the AHC in CGT/5-nm-Pt heterostructure to be 328.4 S/m, which is larger than the experimentally measured value of 82.3 S/m at 4 K. The discrepancy can be caused by several possibilities. First, the CGT/Pt interface in real device is by no means as perfect as that assumed in the calculations, which can reduce the magnitude of AHC. Second, the spin current contribution could produce an AHE signal with the opposite sign. Nevertheless, the good agreement in both sign and magnitude of the AHC for CGT/5 nm-Pt suggests that the induced ferromagnetism in Pt may be the main mechanism of the observed AHE.

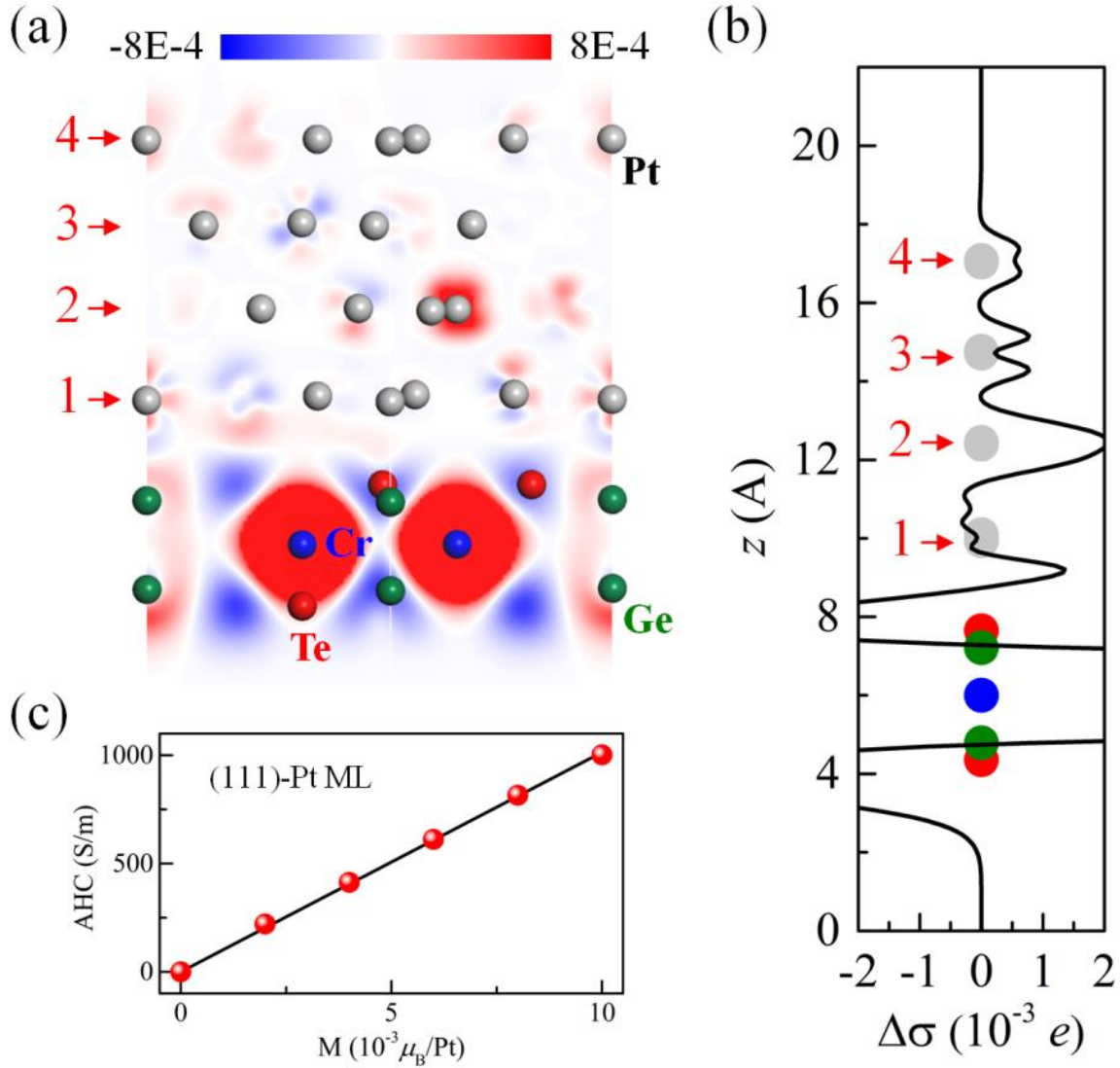

**Figure 5.** (a) Real-space distribution of the spin density difference Δ$\sigma$ and (b) Planar-averaged spin density difference Δ$\sigma$ in CGT/4L-Pt. In (a) and (b), Cr, Ge, Te and Pt atoms are represented by the blue, green, red and gray balls, respectively. The layer index of Pt layer is shown by the numbers in red. (c) Dependence of the AHC on the magnetic moments of Pt in Pt ML.

In conclusion, we have successfully measured the AHE in Pt by forming heterostructures of Pt with thin exfoliated flakes of CGT. The induced AHE as a function of temperature resembles the magnetization of CGT and the distinct features in the AHE hysteresis loops can be mapped to different magnetic domain states in CGT imaged by MFM. DFT calculations show that the observed induced AHE is consistent with the induced moment in Pt arising from the hybridization between Cr 3d electrons and Pt.

Device nanofabrication, transport measurements, and data analyses were supported by DOE BES Award No. DE-FG02-07ER46351. Construction of the pickup-transfer optical microscope and device characterization were supported by NSF-ECCS under Awards No. 1202559 NSF-ECCS and No. 1610447. Theoretical work was supported by DOE Award No: DE-FG02-05ER46237 and calculations were performed on parallel computers at NERSC.

**Methods:**

**Device fabrication process**

We exfoliated the CGT flakes in the load lock of the sputtering system followed immediately by evacuation. Once the load lock pressure reached below $5*10^{-6}$ torr, the samples were loaded into the main chamber which has a base pressure of $10^{-7}$ torr. We next heat the samples in chamber at above 100 ºC to remove water vapor which may have accumulated on the materials surface. Since the oxidation can take place on a short time scale, the surface layer oxidation is unavoidable. To remove the likely oxidized surface layer, we etch the CGT flakes in the sputtering chamber with argon plasma at a power of 15 W with pressure of 40 mtorr, immediately followed by deposition of 5 nm Pt.

**AHE Measurement details**

The transport measurements were performed in a physical properties measurement system by Quantum Design at temperatures down to 4 K. A current of 2 mA is fixed in the device while the potential drop between source and drain is monitored with a Keithley 2400 sourcemeter. Two Keithley 2182A nanovoltmeters are used to monitor $V_{xx}$ and $V_H$. The measurement is setup with the detection direction of $R_H$ determined by the "right-hand rule" to properly determine the sign of the AHC.

**MFM Measurement details**

The MFM measurements were performed in a home-built low temperature scanning probe microscope using commercial MFM probes (Bruker MESP-V2) with a spring constant of

~3 N/m, a resonance frequency at ~ 75 kHz, and a Co-Cr magnetic coating. MFM images were taken in a constant height mode with the tip scanning plane at ~80 nm above the sample surface. The MFM signal, the change in the resonance frequency, is measured by a Nanonis SPM Controller using a phase-lock loop.

**DFT Computational details**

Our DFT calculations are carried out by using VASP[25,26]. Electronic exchange-correlation is described by the generalized-gradient approximation with the functional proposed by Perdew, Berke, and Ernzerhof (PBE)[27]. We utilize projector-augmented wave pseudopotentials to describe core-valence interaction[28,29] and set the energy cutoff for plane-wave expansions to be 500 eV[27]. Atomic structures are fully optimized with a criterion that requires the force on each atom being less than 0.01 eV/Å. The LSDA+U method[30], with an effective $U_{eff}$=1.0 eV[1], is employed to take the correlation effect of Cr $3d$ electrons into account. We include the nonlocal vdW functional (optB86b-vdW)[31,32] to correctly describe the interaction across CGT and Pt layers.

Considering that CGT is a van der Waals material, a CGT monolayer is utilized in building up CGT/Pt heterostructures to reduce computational loads. Moreover, we use $\sqrt{7}\times\sqrt{7}$ supercell of (111)-Pt (lattice constant $a$=7.33 Å) and stretch the lattice constant of CGT monolayer to match this supercell. Two heterostructures, one with 4-layer Pt (CGT/4L-Pt) and the other with 5 nm (22-layer) Pt (CGT/5 nm-Pt), are explicitly considered in this work. Note that we only calculate the former's AHC, since it is tractable to the current computation ability. A 13-Å vacuum layer is adopted to avoid artificial interactions between periodic slabs. We consider three representative atom alignments, namely, one of Pt atoms of the first-layer Pt sitting directly on the top of (i) Ge, (ii) Te and (iii) Cr, respectively (Figure S1). We find that the case (i) is most stable with a much lower energy than cases (ii) and (iii) (Figure S1). So, we focus on the AHC of case (i).